# Tracking the Picoscale Spatial Motion of Atomic Columns During Dynamic Structural Change

Barnaby D.A. Levin, Ethan L. Lawrence & Peter A. Crozier*

*Corresponding Author. Email: crozier@asu.edu

**Abstract**

In many materials systems, such as catalytic nanoparticles, the ability to characterize dynamic atomic structural changes is important for developing a more fundamental understanding of functionality. Recent developments in direct electron detection now allow image series to be acquired at frame rates on the order of 1000 frames per second in bright-field transmission electron microscopy (BF TEM), which could potentially allow dynamic changes in the atomic structure of individual nanoparticles to be characterized with millisecond temporal resolution in favourable cases. However, extracting such data from TEM image series requires the development of computational methods that can be applied to very large datasets and are robust in the presence of noise and in the non-ideal imaging conditions of some types of environmental TEM experiments. Here, we present a two-dimensional Gaussian fitting algorithm to track the position and intensities of atomic columns in temporally resolved BF TEM image series. We have tested our algorithm on experimental image series of Ce atomic columns near the surface of a ceria ($CeO_2$) nanoparticle with electron beam doses of ~125-5000 $e^-Å^{-2}$ per frame. The accuracy of the algorithm for locating atomic column positions is compared to that of the more traditional centroid fitting technique, and the accuracy of intensity measurements is evaluated as a function of dose per frame. The code developed here, and the methodology used to explore the errors and limitations of the measurements, could be applied more broadly to any temporally resolved TEM image series to track dynamic atomic column motion.

1. **Introduction**

Aberration-corrected transmission electron microscopy (TEM) is a powerful tool for characterizing atomic structures with sub-angstrom spatial resolution [1-3]. Atomic resolution TEM images generally have an acquisition time on the order of seconds, and analysis of the images typically treats atomic structures as static. However, in many systems, such as catalytic nanoparticles, the atomic structure may undergo dynamic changes, particularly at the particle surface [3-7]. The ability to characterize these dynamic atomic structural changes is important for developing a more fundamental understanding of catalytic functionality and requires imaging techniques with both atomic-scale spatial resolution and improved temporal resolution.

Recent advances in direct detection technology now allow TEM image series to be acquired at frame rates on the order of 1000 frames per second (fps), with high sensitivity at low electron fluences, or low electron doses, (using terminology more common in the electron microscopy community) [8-13]. This technology has the potential to allow atomic structures to be characterized with millisecond temporal resolution, which can potentially reveal new information about atomic structural dynamics. To extract quantitative information about structural dynamics from temporally resolved image series of a nanoparticle, computational methods are needed to calculate atomic column positions at each point in time. Ideally, these computational methods should also recover information about the three-dimensional structure of the particle. Conventional techniques for three-dimensional reconstruction in TEM such as electron tomography and exit-wave reconstruction are not currently practical for temporally resolved analysis at the millisecond level, because these techniques require a series of a number of images to be taken for each 3D reconstruction [3,14]. In principle, partial 3D information about the sample can be recovered from a single TEM image by measuring the intensity of each atomic column and comparing these results with simulated images to calculate the number of atoms in each column. This has been demonstrated for high-quality TEM images [15], although the limitations of this method for noisier images acquired under low dose conditions require exploration.

There are several key challenges for developing computational methods to identify atomic column positions and intensities in TEM image series. Firstly, such methods must be robust in the presence of noise and non-ideal imaging conditions. Pushing the limits of temporal resolution will necessarily involve a low electron dose in each frame of an image series. For a given incident electron beam intensity, each frame of an image series acquired at 1000 fps will receive only $1/1000^{th}$ of the total electron dose of a single frame acquired over 1 second. This means that the images in each frame of a temporally resolved image series will be significantly noisier than a single frame with 1 second exposure acquired under the same conditions. **The electron dose per frame will therefore be a critical factor limiting our ability to extract quantitative information from temporally resolved image series.** A primary motivation for our work is extracting information from *in situ* datasets. Dynamic changes during *in situ* observation often occur unexpectedly and imaging conditions may not be ideal. For example, nanoparticles often tilt and undergo drift during observation. The sample thickness and defocus may not be optimum when a dynamic event occurs. However, such datasets still contain a wealth of information on atomic-level sample dynamics and robust processing methods should be able to handle these situations.

Secondly, computational methods used to analyze temporally resolved TEM image series must be able to reliably separate signals from background. Achieving a high temporal resolution in TEM is more practical using the bright-field (BF) imaging mode, due to greater signal collection efficiency relative to dark-field imaging. However, BF-TEM images have non-zero intensity in vacuum and local intensity in the sample is either higher or lower than the vacuum intensity (background) due interference effects. With a relatively low electron dose in each frame, some features of interest may appear weak relative to the background and will be challenging to distinguish from noise. This implies that methods developed for locating atomic column positions in annular dark-field scanning TEM (ADF STEM) [16,17], in which the contrast of atomic columns is typically stronger relative to background intensity (at least for heavier elements), may not be robust when applied to temporally resolved BF TEM datasets.

Finally, computational methods used to process and analyze temporally resolved TEM image series must be able to operate on very large datasets. For example, a TEM image series acquired at 1000 fps, on a 4096 x 4096 pixel direct electron detector with data saved in 32-bit format will reach 1 TB in size after only ~15 seconds of total acquisition time. For fast processing of data, simple computational methods are therefore preferable to more complex, computationally demanding methods.

Previous techniques for tracking objects in TEM image series have generally used a centroid (or centre of mass) fitting method [18,19], and this method has recently been applied to temporally resolved series of BF-TEM images to track the positions of atomic columns in a nanoparticle with a precision of ~10 pm [20]. However, a disadvantage of centroid fitting is that there is not a simple way to use this method to separate signal from background and measure atomic column intensities in BF TEM images. As discussed above, recording intensity as well as position is necessary to recover some information about three-dimensional atomic structure from the image series, which is important for the interpretation of phenomena such as dynamic atomic restructuring during catalysis. An alternative to fitting a centroid is to fit a two-dimensional (2D) elliptical Gaussian to each atomic column. This method has previously been used to locate atomic columns in static

TEM and STEM images with picoscale precision and can allow column intensity to be calculated [3,21-27]. Recently developed algorithms for ADF STEM have explored more advanced computational techniques such as machine learning [16], or advanced statistical analyses [25] for atomic column tracking. In these algorithms, a 2D Gaussian is often used as the underlying model to fit to atomic columns. In order for similar techniques to be applied to BF TEM image series, the limitations of the performance of the 2D Gaussian model on noisy BF TEM images needs to be understood.

Here, we present a 2D elliptical Gaussian fitting method, implemented in MATLAB, for simultaneously extracting atomic column positions with sub-pixel precision and extracting atomic column intensities from temporally resolved BF TEM image series, allowing changes of the position and intensity of each column on a bright background to be tracked over time. The method was tested by measuring the picoscale changes of Ce atomic column positions and intensities close to the surface of a ceria ($CeO_2$) nanoparticle in an experimental TEM image series. The performance of the Gaussian fitting algorithm for determining positions is compared to that of the centroid fitting method. The relationship between electron beam dose per frame and the error on intensity measurements is explored. Some simple filtering techniques to improve signal-to-noise without sacrificing temporal resolution are evaluated.

The methodology that we have developed here may be useful for the wider TEM community as advances in instrumentation make tracking atomic structural dynamics more practical. We have therefore made our code, named "Temporally Resolved Atomic Column Tracking" (TRACT) available to download via Github (https://github.com/bdalevin).

## 2. Materials and Methods

2.1 Acquisition of Test Data

Experimental temporally resolved image series of $CeO_2$ nanoparticles imaged in a [110] projection were used to test the algorithms developed for this manuscript. Images were acquired using an aberration-corrected FEI Titan environmental transmission electron microscope, operated at 300 kV. Nanoparticles were observed at room temperature, with a pressure of $<10^{-6}$ Torr at the sample. Negative spherical aberration (Cs) imaging (NCSI) [28] was used for our experiments, with Cs = ~ -20 μm. This resulted in bright contrast for both cerium columns, and the more weakly scattering oxygen atomic columns. An electron dose of 5000 $e^-Å^{-2}s^{-1}$ was applied to the nanoparticle to minimize electron beam damage effects [29,30]. Images were acquired using a Gatan K2 IS direct electron detector, operated at 10-40 frames per second yielding an electron dose of ~125-500 $e^-Å^{-2}$ per frame. The detector was operated in the electron counting mode [31], which reduces the maximum frame rate to ~40 frames per second, but improves signal to noise. The scale of the images acquired was ~4 pm per pixel. In this initial study, we have focused on the cerium cation columns in the image, rather than the oxygen anion columns, which have a lower intensity than the cerium columns, and can be more challenging to detect in individual frames.

2.2 Pre-processing of Data to Reduce Noise

Temporally resolved image series were first aligned to correct for sample drift by cross-correlation in Digital Micrograph (Gatan, Inc.). The data was then exported for processing and analysis in other software. Figure 1a shows an image of an area close to a (111) surface of the particle (a Fourier transform an image of a larger area of the particle is shown in Figure 1b). The image is a sum of multiple individual frames with a total exposure time of 1 second. Figure 1c shows an image of the area indicated by the red box in Figure 1a, but with a total exposure time of only 0.1s. Due to the high levels of noise in individual frames (such as Figure 1c), we chose to test three spatial imaging filters an effort to improve the signal-to-noise ratio in each frame without degrading the temporal resolution of the image series, and without significantly degrading spatial resolution. The three filters tested were the Wiener filter [32,33], TV-L1 filter [34] and the Gaussian blur filter. Of these three, we ultimately chose to use the Gaussian blur filter because it appeared to introduce the fewest image artifacts. Figure 1d shows the result of applying a Gaussian blur to the data in Figure 1c. For brevity, we have chosen to give a more detailed discussion of the effects of the Wiener and TV-L1 filters in the supplementary information, but one particularly important observation to note is that the Wiener filter imposes a periodicity on an image, causing a shift in the apparent position of off-lattice columns such as the column indicated by the white arrow in Figure 1a and Figure 1c (see Supplementary Figures 1-2 and Supplementary Table 1). This suggests that the Wiener filter is not appropriate for the study of dynamic shifts of atomic columns away from lattice sites.

When applying the Gaussian blur filter (implemented in ImageJ) to our data, we use a circular Gaussian with a width that is much narrower than the typical width of an atomic column in our images. For the data presented in this manuscript, we chose a radius of 2 for the Gaussian blur, which corresponds to a full width at half maximum (FWHM) of ~4.7 pixels. A typical atomic column in the data has a FWHM of ~12-20 pixels. We have investigated whether applying the Gaussian blur filter in this way may alter the results of the Gaussian fit by imposing a Gaussian shape on the atomic columns. Supplementary Table 1 compares the $R^2$ metric when fitting Gaussians to columns in the raw data and Gaussian blurred data. $R^2$ values are higher after applying the Gaussian blur, indicating that the Gaussian blur filter does make fitting 2D Gaussians easier.

However, the $R^2$ values are similarly improved when using a non-Gaussian filter such as a Wiener filter or TV filter. This suggests that much of the improvement in fitting may be due to smoothing of the data, rather than the imposition of a Gaussian shape on the atomic columns.

2.3 2D Gaussian Fitting Method

Our atom tracking method, TRACT, works by fitting 2D elliptical Gaussians to TEM image series containing atomic-resolution images using least squares curve fitting in MATLAB. 2D Gaussians are fitted about a user-specified array of points in the image, which should roughly correspond to the positions of atomic columns in the image. The general formula used for 2D Gaussian fitting is given by Equation 1.

**Equation 1**

$$F(x,y) = G(x,y) + B$$

where:

$$G(x,y) = A e^{-\frac{1}{2}\left(\frac{(x-x_0)\cos\theta - (y-y_0)\sin\theta}{\sigma_a}\right)^2} e^{-\frac{1}{2}\left(\frac{(x-x_0)\sin\theta + (y-y_0)\cos\theta}{\sigma_b}\right)^2}$$

There are 7 fit parameters in total. $A$ is the amplitude of the Gaussian. $x_0$, $y_0$ are the coordinates of the maximum, which can be determined with sub-pixel precision, and which we define as the position of the atomic column. $\sigma_a^2$ and $\sigma_b^2$ are the variances of the Gaussian along the major and minor axes. $\theta$ is the angle of rotation of the major axis from the x-axis of the image. In BF-TEM imaging the detector will register some spatially non-varying (or slowly varying) signal even when no atomic columns are present. The constant term $B$ in Equation 1 accounts for this spatially non-varying contribution, which we refer to as the background. This allows us to separate the signal attributable to the atomic column, $G(x,y)$, from the signal attributable to the bright background.

2.4 Initialization and Implementation for Fitting Gaussians to Multiple Atomic Columns in an Image Series

After loading image series into MATLAB, approximate positions for each atomic column of interest are generated and used as initial guesses for the Gaussian fitting algorithm. For large datasets, this is achieved by using a MATLAB script to generate a periodic array of positions based on the periodicity of atomic columns in the image. The user then removes excess positions from this array so that the set of initial positions to be fed into the algorithm corresponds only to positions in the image where the user expects to see (or suspects there may be) atomic columns. The user may also add positions to the initial array manually, which can be helpful for defect sites that may be far from a lattice site. Once the array of initial guesses for column positions has been finalized, an identification number is assigned to each of the atomic columns, allowing changes in their position and intensity to be tracked between different frames in the image series. Initial guesses and upper and lower bounds for each of the fit parameters, including amplitude, variance, and background, are chosen based on an examination of the size and intensity of atomic columns and background in the images. The upper and lower bounds placed on the fitting parameters act as constraints to ensure that the Gaussians are fitted to cerium columns and are not mistakenly deflected onto oxygen columns. To ensure a more accurate set of initial parameters for Gaussian fitting to noisy data in individual frames, the algorithm is initially run on the sum of all of the images in the image series. The position of each atomic column in the summed image series is precisely located by first finding the local maximum within a user-specified radius from each initial position, fitting a centroid around each local maximum (limited to a small radius of ~10 pixels), and then fitting a 2D elliptical Gaussian around each centroid position. The coordinates of the maxima of each of the 2D Gaussians in the summed image define the positions of each atomic column. The centroid and Gaussian fitting stages of coding both make use of the particle tracking scripts originally written for the programming language IDL by Crocker, Grier, and Weeks [18] and implemented in MATLAB by Blair & Dufresne (http://physics.georgetown.edu/matlab/).

It is important to note that our approach to initialization as described above has been tailored to our specific experiment, where we aim to track relatively small displacements of cation columns. For experiments of a different nature to the example presented in this manuscript, for example experiments in which rapid dynamic events preclude clear identification of all atomic columns in the summed image, alternative methods may be more practical for identifying an initial set of atomic column positions. Various methods described in the literature include those based on non-local means and adaptive periodic searching [35], blob detection [36], and pattern matching [37]. An alternative initialization method that we attempted to apply to our data was to find atomic column positions by identifying local maxima in each frame of the image series, subject to different thresholding conditions. The attraction of this alternative was that it was independent of the periodicity of the structure, but unfortunately high levels of noise in our data meant that there were too many false positives for the local maxima method to be practical for our purposes.

After atomic column positions are found in the summed image, a 2D Gaussian fit is applied to each atomic column in each individual frame of the image series, using the parameters obtained from fitting to the summed image as initial inputs. In an effort to ensure that Gaussians are fitted to genuine atomic columns, rather than noise, fit parameters were only saved and used for analysis if the amplitude of the Gaussian exceeds a threshold. We set this threshold at 2 times the level of the noise in the image, which would correspond to a 95% confidence level assuming normal statistics.

2.5 Extracting Atomic Column Positions, Intensities and Associated Errors

Once Gaussians were fitted to all atomic columns in an image series, the position of each column, $x_0, y_0$ may be extracted, and changes in the positions of the columns over time may be analyzed.

The error on the measured position of the atomic columns may be taken by measuring the standard deviation of the measurements of the position of an object that is expected to remain stationary, as has been done in previous work involving centroid fitting [20]. We also note studies that have derived a simple expression for the theoretical best precision that can be obtained in measuring the position of an object as being approximately equal to the width of the object (for atomic columns imaged under ideal conditions, this should approximately equal the resolution limit of the microscope) divided by the square root of the beam dose incident on the object in a given frame [38,39]. This formula is useful for comparison to experimentally derived estimates of positional error.

The intensity attributable to each atomic column is calculated as the intensity of the 2D Gaussian, $G(x,y)$, with the constant term, $B$, treated as background. To aid the reader in following the calculations, it is helpful to perform a simple rotation of the coordinate axes to remove the angular term $\theta$. The 2D Gaussian formula then simplifies to:

**Equation 2**

$$G(x', y') = Ae^{-\frac{1}{2}\left(\frac{x'-x'_0}{\sigma_a}\right)^2} e^{-\frac{1}{2}\left(\frac{y'-y'_0}{\sigma_b}\right)^2}$$

In this form, a Gaussian may be integrated analytically between ± ∞. In one dimension, this gives:

$$Column\ Intensity\ (\pm \infty, 1D) = A \int_{-\infty}^{+\infty} e^{-a(x+c1)^2} dx = A\sqrt{\frac{\pi}{a}}$$

In two dimensions, this gives:

$$Intensity\ (\pm \infty, 2D) = A \int_{-\infty}^{+\infty} e^{-a(x+c1)^2} dx \int_{-\infty}^{+\infty} e^{-b(y+c2)^2} dy = A\sqrt{\frac{\pi}{a}}\sqrt{\frac{\pi}{b}} = \frac{A\pi}{\sqrt{ab}}$$

In our case, $a = \frac{1}{2\sigma_a^2}$, and $b = \frac{1}{2\sigma_b^2}$. This implies that:

$$Column\ Intensity\ (\pm \infty) = \frac{A\pi}{\sqrt{ab}} = 2A\pi\sigma_a\sigma_b$$

One method of estimating the error on intensity is to take the standard deviation of the intensities of a column that one expects to have a constant occupancy as described above for estimating the error on positions. A second method to calculate errors in column intensity due to noise is to do so on a column by column basis. To calculate the error on intensities in this manner, we assume that the variance of the total signal, and the variance of the background follow Poisson statistics, which may be reasonable for a direct detector in counting mode since other sources of noise such as Landau noise are eliminated (Li et. al. 2013). It is important to define an area over which to measure the noise for this method to be successful. Bounds of ± ∞ are impractical, so instead we choose to define this area as an ellipse with major and minor radii $2\sigma_a$ and $2\sigma_b$ (60 – 80). To ensure that noise and intensity are calculated within the same area, we also scale the calculated intensity of the Gaussian. Numerically, in one dimension, 95.45% of a Gaussian's total integrated intensity lies within a radius of 2σ. In 2 dimensions, this is (95.45%)$^2$ or 91.11% of the Gaussian's total integrated intensity.

Our formula for calculating atomic column intensity from our Gaussian fits is therefore:

**Equation 3:**

$$Column\ Intensity\ (\pm 2\sigma) = 0.9111 \times 2A\pi\sigma_a\sigma_b$$

The variance, extracted column intensity may be derived as follows:

**Equation 4:**

i. $I_{Total} = I_{Column} + I_{Background}$

ii. $Variance(I_{Total}) = I_{Total}$

iii. $Variance(I_{Background}) = I_{Background}$

iv. $Variance(I_{Column}) = Variance(I_{Total}) + Variance(I_{Background})$

v. $Variance(I_{Column}) = I_{Column} + 2I_{Background}$

The error is given by the square root of the variance within our ellipse defined by radii $2\sigma_a$ and $2\sigma_b$, with area $4\pi\sigma_a\sigma_b$. This implies that error is given by:

**Equation 5:**

$$Column\ Intensity\ Error \approx \sqrt{Column\ Intensity + 8B\pi\sigma_a\sigma_b}$$

More rigorous treatments of errors on intensity due to noise have been applied to STEM imaging [40]. Adapting such treatments to our BF-TEM data is an area of future interest but is beyond the scope of the current manuscript and the simpler treatment outlined above should provide a reasonable approximation.

2.6 Estimating Atomic Column Occupancy from Gaussian Intensity

A limited amount of information about the three-dimensional structure of a TEM specimen can in principle be recovered by comparing the intensities of atomic columns in experimental data with those in simulated images [15]. Multislice simulations of a material surface can be performed with varying atomic column occupancies. The intensities of the atomic columns in the simulated images can be measured using the Gaussian fitting procedure described above, and then compared to intensities measured in experimental data. This can allow the number of atoms in each column in the experimental data to be estimated. The accuracy of this estimate will be limited by the contrast of the atomic columns, and the level of noise of the image. The output of the simulated data may have different units, and a different pixel density to the experimental data. In order to perform a comparison, both the simulated and experimental images should be normalized such that the intensity in a given area of vacuum (far from the sample) are equal in both sets of data. We have chosen to normalize the data such that the integrated intensity in an area of vacuum equal to 1 Å$^2$ is equal to 1. In our experiments, this normalization was performed by averaging over a large area in vacuum (> 100 Å$^2$), and so the error associated with the normalization is expected to be much smaller than the error associated with Equation 4. Simulations and comparison between data and experiment are described in the Results and Discussion section below.

## 3. Results and Discussion

### 3.1 Measuring Changes in Atomic Column Position

The most fundamental output of the TRACT code is a data file containing the 2D Gaussian fit parameters of every atomic column analyzed, in every frame of an image series. This includes the x and y coordinates of each column in each frame, along with the intensities of those columns, and an estimate of the error of the intensity due to noise. With the positions and intensities of each of the atomic columns known over time, one can perform more complex analysis of the data, by tracking the motion of atomic columns over time, and modelling changes in column occupancy with time. Some basic examples of each of these analyses are given below. Figure 3 shows the results of a measurement of four Ce atomic column positions in two different frames on the same area of a $CeO_2$ nanoparticle (111) surface. The measured difference in position for three of the four atomic columns are found to be 2 pm or less, which is less than half of the width of 1 pixel in the data. However, for the top left column, a difference in position of 8.5 pm is measured, which is over twice the width of a pixel in the data.

Figure 3 shows an evaluation of the differences in position between four atomic columns in two frames, but a more typical temporally resolved image series will likely require analysis of many atomic column positions over many frames. This makes comparison between positions in individual frames laborious, and therefore simpler methods of comparison are required. One technique that is useful for quickly identifying the atomic columns that exhibit the greatest degree of motion is to plot the standard deviation of the atomic column position, as shown in Figure 4a below. The Figure shows a more extended area of the (111) $CeO_2$ nanoparticle surface from Figure 3. Coloured circles overlaid on the atomic columns indicate the magnitude of the standard deviation in atomic column position. The atomic columns with the largest standard deviation in position are generally located on the surface of the particle, particularly at step-edge sites.

Calculating standard deviations is also useful for obtaining an estimate in the error on the measurement of column position. For the $CeO_2$ nanoparticle in Figure 4, atomic columns away from the surface are not expected to move significantly because bulk oxygen ion transport is very slow at room temperature in $CeO_2$, and atomic columns away from the surface cannot interact with the environment. Assuming that the motion of Ce atoms on the top and bottom surfaces of a column have only a small effect on the apparent motion of the whole column, measured displacements of sub-surface atomic columns between frames may be assumed to result from statistical errors due to noise. The typical standard deviation of sub-surface column positions (~5 pm) can be used to characterize the error of the measurement of position. This rationale has also been applied in previous work using centroid fitting [20]. Assuming normal statistics, a measured displacement of greater than 4.7 pm can be considered genuine at the 68% confidence level, and a displacement of greater than 9.4 pm can be considered genuine at the 95% confidence level. For comparison, we repeated the measurement of the standard deviation in column positions using the centroid fitting method (Figure 4b). The results are qualitatively similar to those obtained by Gaussian fitting in that they show a larger standard deviation in position for surface columns than for sub-surface columns. However, the typical standard deviation of a sub-surface column is larger for centroid fitting (~7.6 pm) than it is for Gaussian fitting, suggesting that the Gaussian fitting algorithm may be more accurate at identifying atomic column positions. Both methods give larger errors on position than the theoretical expression for the best precision [38,39], which gives ~2.6 pm. The standard deviation of position of sub-surface columns is explored as a function of dose per frame in Figure 4c and Supplementary Table 2. It is possible that for our dataset, the centroid fits did not perform as well as the Gaussian fits because they were more easily skewed by fluctuations in noise, or by the intensity of oxygen columns, adjacent to the Ce columns. The standard deviation measured by centroid fitting also appears to be lower for some atomic columns at the surface than that measured by Gaussian fitting. This may result from the fact that because the Gaussian fitting algorithm is a parametric model-based method, we were able to exclude frames from the calculation in which the parameters of the fitted Gaussian suggested no atomic column appeared to be present, as described in the methods section above. As a simple non-parametric weighted average method, the centroid fitting algorithm is not able to perform a similar exclusion, so the standard deviation of the position of surface columns may be skewed by centroid measurements of frames in which an atomic column is not clearly visible above the noise level.

3.2 Estimating Atomic Column Occupancy

Multislice simulations for $CeO_2$ were performed using the JEMS software package (Stadelmann, http://www.jems-saas.ch/) Imaging conditions for the simulations were carefully chosen to match those of the experimental data, with the optic axis along the [110] direction. Parameters for the simulations are given in Supplementary Information. The effects of defocus and tilt are important to account for when performing image simulations, and particularly when comparing simulated column intensities to experimental intensities. For this dataset, the defocus is approximately constant over the 1 s exposure time of the experiment. For the current analysis, we concentrate only on atomic columns located close to the surface where the sample is thin. Under these conditions, tilt effect are minimal. Thicknesses of 1-12 Cerium atoms per column were simulated. It was noted that the intensity of Ce columns increased from 1 to 6 atoms per column, and then began to decrease from 7 to 12 atoms per column. Oxygen column intensities were not measured, but it was noted that these appeared to increase as the number of atoms per column increased from 1-12 atoms and did not appear to decrease. The intensity of the Ce columns above the background was measured by Gaussian fitting at each thickness using Equation 3. This allowed a lookup table to be generated indicating the expected intensity of Ce columns with a given occupancy. In order for the lookup table and experimental data to be comparable, both were normalized such that the integrated intensity of an area of 1 $Å^2$ in vacuum was equal to 1, as described in Methods above. Error bars in the lookup table indicate the range of measured intensities between Ce columns at surface and sub-surface sites in the simulated structure, as well as sites where the Ce atoms are located in different depths along the [110] axis. Results from the simulations are shown in Figure 5. In our experimental data from a $CeO_2$ nanoparticle, the intensity of oxygen columns at the (111) surface is very low. This suggests that the thickness of the particle at this surface is in the 1-6 atom per column range. From the experimental data, atomic column intensity can be measured by Gaussian fitting using Equation 3. As stated in methods above, we have considered two methods for estimating the error on the intensity measurement. Here, we first investigate the error calculated by estimating Poisson Noise (Equation 4). In order to investigate the relationship between electron dose and the error in intensity measurement, the results of the intensity of an atomic column is analyzed as a function of dose per frame in Figure 6. These measurements suggest that, when using the experimental conditions described in Methods, an electron dose of at > 2500 $e^- Å^{-2}$ per frame is needed to determine that the occupancy of an individual column of 3 Ce atoms to accuracy of less than ± 1 atom at 95% confidence. Using the same method, we can determine that in Figure 6b, there are 2 atoms in the top right column and 3 atoms in each of the others to 95% confidence. It should be noted that even at 5000 $e^- Å^{-2}$ per frame, the 95% confidence level error bars exceed the difference in intensity between 5 and 6 atoms per column, implying that a dose much greater than 5000 e$^-$Å$^{-2}$ per frame may be required to determine these occupancies accurately.

We have also considered an alternative method of calculating the error on intensity based on the standard deviation of a series of measurements of the intensity of an atomic column with a constant number of atoms per column. We find that this method gives an error that is slightly larger than that given by the Poisson noise analysis, but that it converges towards the value given by the Poisson noise analysis as the dose per frame is increased. This is discussed in greater detail in the Supplementary Information (see Supplementary Figure 3 and associated discussion).

In general, using a larger electron beam dose is likely to lead to improved precision of intensity measurement. However, increasing electron dose will also increase the probability that any dynamic changes observed are caused by electron beam damage, or that electron beam damage may alter the structure *during* the acquisition of an image. This may be undesirable if the goal of an experiment is to observe only non-beam-induced processes, although in some situations, it may be desirable to use the effects of the electron beam to drive physical changes in the specimen. It is possible that varying microscope parameters such as electron beam voltage and defocus may improve the contrast of the atomic columns and allow column occupancy to be determined with greater accuracy at lower beam dose. However, extensive multislice simulations would be needed to determine optimum conditions. The optimum conditions are likely to differ for different materials and may differ when imaging the material at different orientations.

## 4. Conclusions/Summary

In summary, we have developed an algorithm to track the position and intensities of atomic columns in temporally resolved BF TEM image series using 2D Gaussian fitting. This algorithm has been tested on Ce atomic columns in images of a $CeO_2$ nanoparticle acquired with a negative spherical aberration imaging technique. We have evaluated several commonly used image filters for pre-processing data, finding that the Gaussian blur filter is reasonable, but that the Wiener Filter is inappropriate, as it reduces the contrast of defects in the sample. The accuracy of the 2D Gaussian fitting algorithm for measuring position was compared with centroid fitting by measuring the standard deviation in the position of stationary sub-surface columns in the $CeO_2$ nanoparticle. The Gaussian fitting method was found to be slightly more accurate than centroid fitting in this case. The accuracy of using intensity measurements to determine atomic column occupancy was evaluated assuming error due to Poisson noise. For our conditions, it was found that an electron beam dose of 2500 $e^-Å^{-2}$ per frame or greater was required to determine Ce column occupancy to 95% confidence. Electron dose per frame is a critical limiting factor in accurately tracking atomic column positions and intensities. In general, a higher electron beam dose per frame should yield improved signal-to-noise, which should in turn yield more precise measurements. However, high electron dose per frame will also increase the probability that artifacts may be introduced into the data due to radiation damage effects. In future related work, we aim to use the methods presented here to analyse the relationship between Ce column motion and atomic site in more detail, and to analyze dynamic restructuring of $CeO_2$ in images acquired with greater dose per frame, and at greater temporal resolution. We believe that the code developed here, and the methodology used to explore the errors and limitations of the measurements, could be applied more broadly to any temporally resolved TEM image series to track dynamic atomic column motions, and have made our MATLAB code available for the community to use on Github (https://github.com/bdalevin). Further work, which will require extensive collaboration with specialists in advanced image processing techniques, may examine the effects of different and more advanced imaging filters for noise reduction, a more sophisticated treatment of the noise related errors on column positions and intensities, and incorporating more advanced computational methods such as machine learning into the 2D Gaussian algorithm.

Acknowledgements

This work was supported by the National Science Foundation (NSF) and Arizona State University. NSF DMR 1308085 supported ELL and PAC as well as data acquisition. NSF DMR 1840841 supported BDAL and PAC as well as coding and data processing. We thank Arizona State University's John M. Cowley Center for High Resolution Electron Microscopy for microscope use. We thank Tara M. Boland and Piyush Haluai of Arizona State University for useful discussions regarding image simulations and experimental noise.

Declaration of Interest Statement

Declarations of interest: None.

Data Statement

The raw experimental data, and the cropped, Gaussian blurred data that was analyzed in this study will be made available via the Figshare repository following acceptance of the manuscript. The files are available in the easily readable 32-bit .tif format.

References

[1] M. Haider, H. Rose, S. Uhlemann, E. Schwan, B. Kabius, K. Urban, A spherical-aberration-corrected 200kV transmission electron microscope, Ultramicroscopy. 75 (1998) 53–60. doi:10.1016/S0304-3991(98)00048-5.

[2] L.C. Gontard, L.-Y. Chang, C.J.D. Hetherington, A.I. Kirkland, D. Ozkaya, R.E. Dunin-Borkowski, Aberration-Corrected Imaging of Active Sites on Industrial Catalyst Nanoparticles, Angew. Chem. Int. Ed. 46 (2007) 3683–3685. doi:10.1002/anie.200604811.

[3] K.W. Urban, Studying Atomic Structures by Aberration-Corrected Transmission Electron Microscopy, Science. 321 (2008) 506–510. doi:10.1126/science.1152800.

[4] P.L. Hansen, Atom-Resolved Imaging of Dynamic Shape Changes in Supported Copper Nanocrystals, Science. 295 (2002) 2053–2055. doi:10.1126/science.1069325.

[5] U.M. Bhatta, I.M. Ross, T.X.T. Sayle, D.C. Sayle, S.C. Parker, D. Reid, S. Seal, A. Kumar, G. Möbus, Cationic Surface Reconstructions on Cerium Oxide Nanocrystals: An Aberration-Corrected HRTEM Study, ACS Nano. 6 (2012) 421–430. doi:10.1021/nn2037576.

[6] S. Takeda, Y. Kuwauchi, H. Yoshida, Environmental transmission electron microscopy for catalyst materials using a spherical aberration corrector, Ultramicroscopy. 151 (2015) 178–190. doi:10.1016/j.ultramic.2014.11.017.

[7] S. Helveg, C.F. Kisielowski, J.R. Jinschek, P. Specht, G. Yuan, H. Frei, Observing gas-catalyst dynamics at atomic resolution and single-atom sensitivity, Micron. 68 (2015) 176–185. doi:10.1016/j.micron.2014.07.009.

[8] A.R. Faruqi, R. Henderson, M. Pryddetch, P. Allport, A. Evans, Direct single electron detection with a CMOS detector for electron microscopy, Nuclear Instruments and Methods in Physics Research Section A: Accelerators, Spectrometers, Detectors and Associated Equipment. 546 (2005) 170–175. doi:10.1016/j.nima.2005.03.023.

[9] R.S. Ruskin, Z. Yu, N. Grigorieff, Quantitative characterization of electron detectors for transmission electron microscopy, Journal of Structural Biology. 184 (2013) 385–393. doi:10.1016/j.jsb.2013.10.016.

[10] G. McMullan, A.R. Faruqi, D. Clare, R. Henderson, Comparison of optimal performance at 300 keV of three direct electron detectors for use in low dose electron microscopy, Ultramicroscopy. 147 (2014) 156–163. doi:10.1016/j.ultramic.2014.08.002.

[11] H.-G. Liao, D. Zherebetskyy, H. Xin, C. Czarnik, P. Ercius, H. Elmlund, M. Pan, L.-W. Wang, H. Zheng, Facet development during platinum nanocube growth, Science. 345 (2014) 916–919. doi:10.1126/science.1253149.

[12] F. Panciera, Y.-C. Chou, M.C. Reuter, D. Zakharov, E.A. Stach, S. Hofmann, F.M. Ross, Synthesis of nanostructures in nanowires using sequential catalyst reactions, Nature Mater. 14 (2015) 820–825. doi:10.1038/nmat4352.

[13] K. Sato, H. Yasuda, Fluctuation of long-range order in Co-Pt alloy nanoparticles revealed by time-resolved electron microscopy, Appl. Phys. Lett. 110 (2017) 153101. doi:10.1063/1.4980077.

[14] B.D.A. Levin, E. Padgett, C.-C. Chen, M.C. Scott, R. Xu, W. Theis, Y. Jiang, Y. Yang, C. Ophus, H. Zhang, D.-H. Ha, D. Wang, Y. Yu, H.D. Abruña, R.D. Robinson, P. Ercius, L.F. Kourkoutis, J. Miao, D.A. Muller, R. Hovden, Nanomaterial datasets to advance tomography in scanning transmission electron microscopy, Sci Data. 3 (2016) 160041. doi:10.1038/sdata.2016.41.

[15] C.L. Jia, S.B. Mi, J. Barthel, D.W. Wang, R.E. Dunin-Borkowski, K.W. Urban, A. Thust, Determination of the 3D shape of a nanoscale crystal with atomic resolution from a single image, Nature Mater. 13 (2014) 1044–1049. doi:10.1038/nmat4087.

[16] M. Ziatdinov, O. Dyck, A. Maksov, X. Li, X. Sang, K. Xiao, R.R. Unocic, R. Vasudevan, S. Jesse, S.V. Kalinin, Deep Learning of Atomically Resolved Scanning Transmission Electron Microscopy Images: Chemical Identification and Tracking Local Transformations, ACS Nano. 11 (2017) 12742–12752. doi:10.1021/acsnano.7b07504.

[17] T. Furnival, D. Knez, E. Schmidt, R.K. Leary, G. Kothleitner, F. Hofer, P.D. Bristowe, P.A. Midgley, Adatom dynamics and the surface reconstruction of Si(110) revealed using time-resolved electron microscopy, Appl. Phys. Lett. 113 (2018) 183104. doi:10.1063/1.5052729.

[18] J.C. Crocker, D.G. Grier, Methods of Digital Video Microscopy for Colloidal Studies, Journal of Colloid and Interface Science. 179 (1996) 298–310. doi:10.1006/jcis.1996.0217.

[19] P.Y. Huang, S. Kurasch, J.S. Alden, A. Shekhawat, A.A. Alemi, P.L. McEuen, J.P. Sethna, U. Kaiser, D.A. Muller, Imaging Atomic Rearrangements in Two-Dimensional Silica Glass: Watching Silica's Dance, Science. 342 (2013) 224–227. doi:10.1126/science.1242248.

[20] Z. Hussaini, P.A. Lin, B. Natarajan, W. Zhu, R. Sharma, Determination of atomic positions from time resolved high resolution transmission electron microscopy images, Ultramicroscopy. 186 (2018) 139–145. doi:10.1016/j.ultramic.2017.12.018.

[21] L. Houben, A. Thust, K. Urban, Atomic-precision determination of the reconstruction of a tilt boundary in by aberration corrected HRTEM, Ultramicroscopy. 106 (2006) 200–214. doi:10.1016/j.ultramic.2005.07.009.

[22] I. MacLaren, R. Villaurrutia, B. Schaffer, L. Houben, A. Peláiz-Barranco, Atomic-Scale Imaging and Quantification of Electrical Polarisation in Incommensurate Antiferroelectric Lanthanum-Doped Lead Zirconate Titanate, Adv. Funct. Mater. 22 (2012) 261–266. doi:10.1002/adfm.201101220.

[23] A.B. Yankovich, B. Berkels, W. Dahmen, P. Binev, S.I. Sanchez, S.A. Bradley, A. Li, I. Szlufarska, P.M. Voyles, Picometre-precision analysis of scanning transmission electron microscopy images of platinum nanocatalysts, Nat Commun. 5 (2014) 4155. doi:10.1038/ncomms5155.

[24] M. Nord, P.E. Vullum, I. MacLaren, T. Tybell, R. Holmestad, Atomap: a new software tool for the automated analysis of atomic resolution images using two-dimensional Gaussian fitting, Adv Struct Chem Imag. 3 (2017) 9. doi:10.1186/s40679-017-0042-5.

[25] J. Fatermans, S. Van Aert, A.J. den Dekker, The maximum a posteriori probability rule for atom column detection from HAADF STEM images, Ultramicroscopy. 201 (2019) 81–91. doi:10.1016/j.ultramic.2019.02.003.

[26] D. Mukherjee, L. Miao, G. Stone, N. Alem, mpfit: a robust method for fitting atomic resolution images with multiple Gaussian peaks, Adv Struct Chem Imag. 6 (2020) 1. https://doi.org/10.1186/s40679-020-0068-y.

[27] C. Ophus, H.I. Rasool, M. Linck, A. Zettl, J. Ciston, Automatic software correction of residual aberrations in reconstructed HRTEM exit waves of crystalline samples, Adv Struct Chem Imag. 2 (2016) 15. https://doi.org/10.1186/s40679-016-0030-1.

[28] C.-L. Jia, M. Lentzen,, K. Urban, High-Resolution Transmission Electron Microscopy Using Negative Spherical Aberration, Microsc Microanal. 10 (2004) 174–184. doi:10.1017/S1431927604040425.

[29] A.C. Johnston-Peck, J.S. DuChene, A.D. Roberts, W.D. Wei, A.A. Herzing, Dose-rate-dependent damage of cerium dioxide in the scanning transmission electron microscope, Ultramicroscopy. 170 (2016) 1–9. https://doi.org/10.1016/j.ultramic.2016.07.002.

[30] R. Sinclair, S.C. Lee, Y. Shi, W.C. Chueh, Structure and chemistry of epitaxial ceria thin films on yttria-stabilized zirconia substrates, studied by high resolution electron microscopy, Ultramicroscopy. 176 (2017) 200–211. https://doi.org/10.1016/j.ultramic.2017.03.015.

[31] X. Li, P. Mooney, S. Zheng, C.R. Booth, M.B. Braunfeld, S. Gubbens, D.A. Agard, Y. Cheng, Electron counting and beam-induced motion correction enable near-atomic-resolution single-particle cryo-EM, Nat Methods. 10 (2013) 584–590. doi:10.1038/nmeth.2472.

[32] R. Kilaas, Optimal and near-optimal filters in high-resolution electron microscopy, Journal of Microscopy. 190 (1998) 45–51. doi:10.1046/j.1365-2818.1998.3070861.x.

[33] L.D. Marks, Wiener-filter enhancement of noisy HREM images, Ultramicroscopy. 62 (1996) 43–52. doi:10.1016/0304-3991(95)00085-2.

[34] L.I. Rudin, S. Osher, E. Fatemi, Nonlinear total variation based noise removal algorithms, Physica D: Nonlinear Phenomena. 60 (1992) 259–268. doi:10.1016/0167-2789(92)90242-F.

[35] N. Mevenkamp, A.B. Yankovich, P.M. Voyles, B. Berkels, Non-local Means for Scanning Transmission Electron Microscopy Images and Poisson Noise based on Adaptive Periodic Similarity Search and Patch Regularization, (n.d.) 8.

[36] B.P. Marsh, N. Chada, R.R. Sanganna Gari, K.P. Sigdel, G.M. King, The Hessian Blob Algorithm: Precise Particle Detection in Atomic Force Microscopy Imagery, Sci Rep. 8 (2018) 978. https://doi.org/10.1038/s41598-018-19379-x.

[37] S. Somnath, C.R. Smith, S.V. Kalinin, M. Chi, A. Borisevich, N. Cross, G. Duscher, S. Jesse, Feature extraction via similarity search: application to atom finding and denoising in electron and scanning probe microscopy imaging, Adv Struct Chem Imag. 4 (2018) 3. https://doi.org/10.1186/s40679-018-0052-y.

[38] S. Van Aert, A.J. den Dekker, D. Van Dyck, A. van den Bos, High-resolution electron microscopy and electron tomography: resolution versus precision, Journal of Structural Biology. 138 (2002) 21–33. https://doi.org/10.1016/S1047-8477(02)00016-3.

[39] S. Van Aert, Model-Based Electron Microscopy, in: P.W. Hawkes, J.C.H. Spence (Eds.), Springer Handbook of Microscopy, Springer International Publishing, Cham, 2019: pp. 2–2. https://doi.org/10.1007/978-3-030-00069-1_12.

[40] A.J. den Dekker, J. Gonnissen, A. De Backer, J. Sijbers, S. Van Aert, Estimation of unknown structure parameters from high-resolution (S)TEM images: What are the limits?, Ultramicroscopy. 134 (2013) 34–43. https://doi.org/10.1016/j.ultramic.2013.05.017.

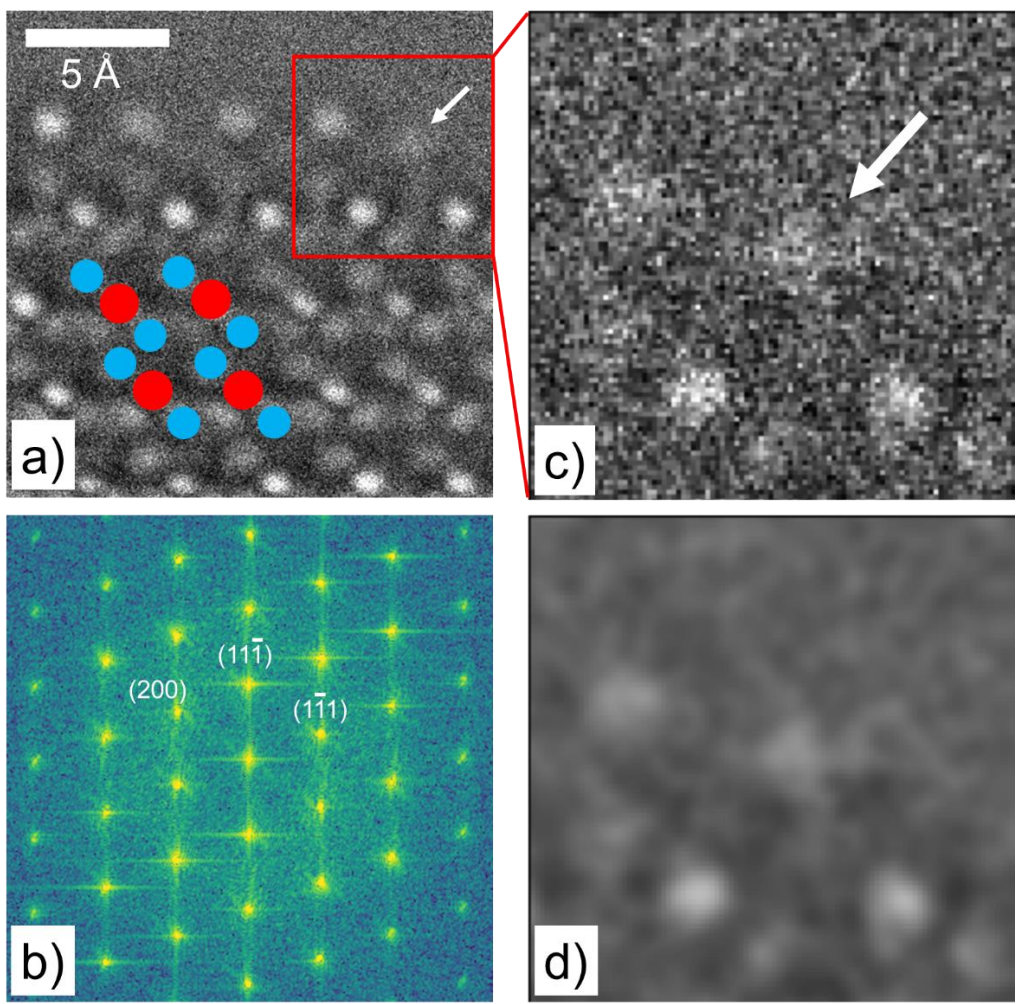

**Figure 1.** a) Unfiltered, summed image of a step edge on a (111) $CeO_2$ nanoparticle surface with a total 1s exposure viewed in a [110] projection. Circles overlaid on the image indicate the projected crystal structure, with red circles representing Ce columns, and blue circles representing O columns. The white arrow indicates an atomic column in an off-lattice (defect) position. b) Central area of a Fourier transform of an image of a larger area of the $CeO_2$ nanoparticle particle c) Unfiltered image with 0.1 s total exposure time of area of the Ce nanoparticle indicated by the red box in (a). The white arrow again highlights a Ce column in an off-lattice position. The electron dose used to acquire this frame was ~500 $e^-Å^{-2}$. The result of applying a Gaussian blur filter of radius 2 pixels to (c).

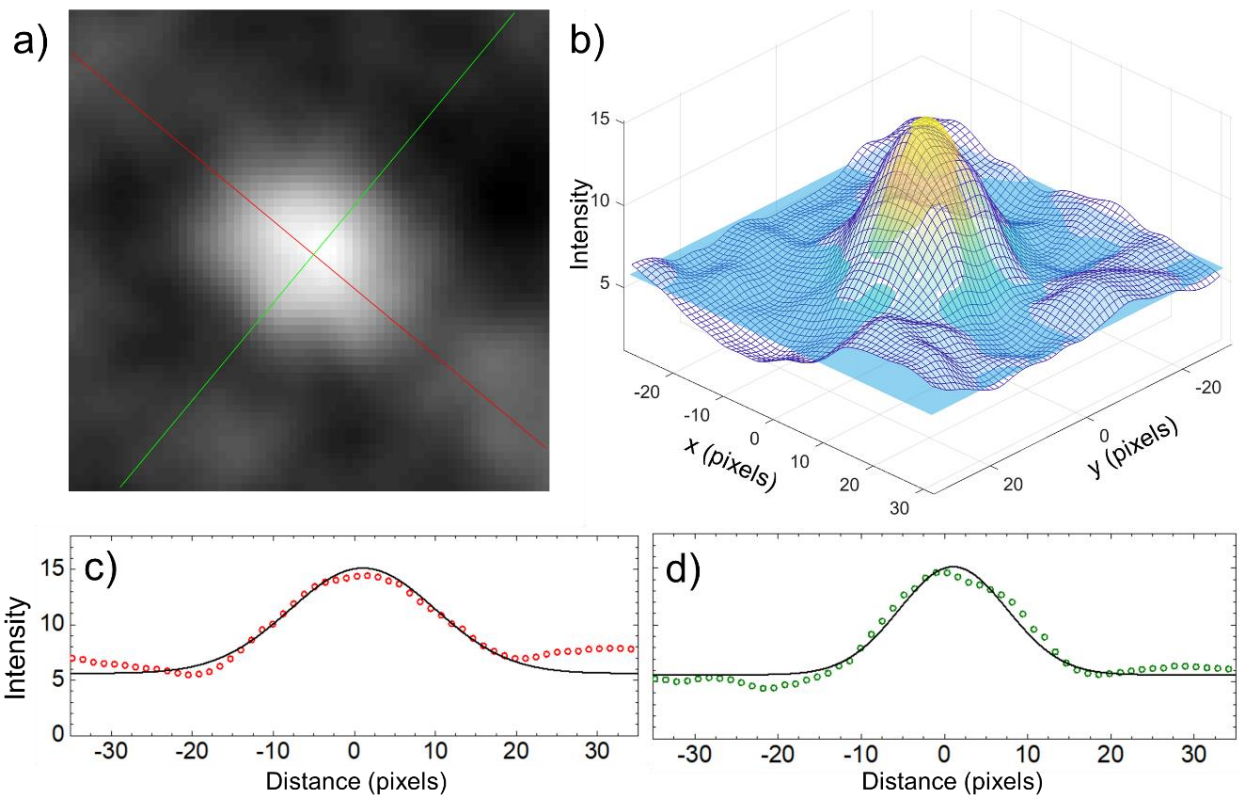

**Figure 2.** a) Image of a Ce atomic column in a $CeO_2$ nanoparticle, with 1 second total exposure, after applying a Gaussian blur. The red and green lines across the image respectively show the major and minor axes of a 2D Gaussian fitted to the column. b) A 3D representation of the 2D Gaussian fit. c) A 1D cross section through the data along the major axis of the 2D Gaussian fit. The red circles represent image intensities, and the black line is the cross section of the fit. d) A 1D cross section through the data along the minor axis of the 2D Gaussian fit. The green circles represent image intensities, and the black line is the cross section of the fit (pixel size = 4 pm).

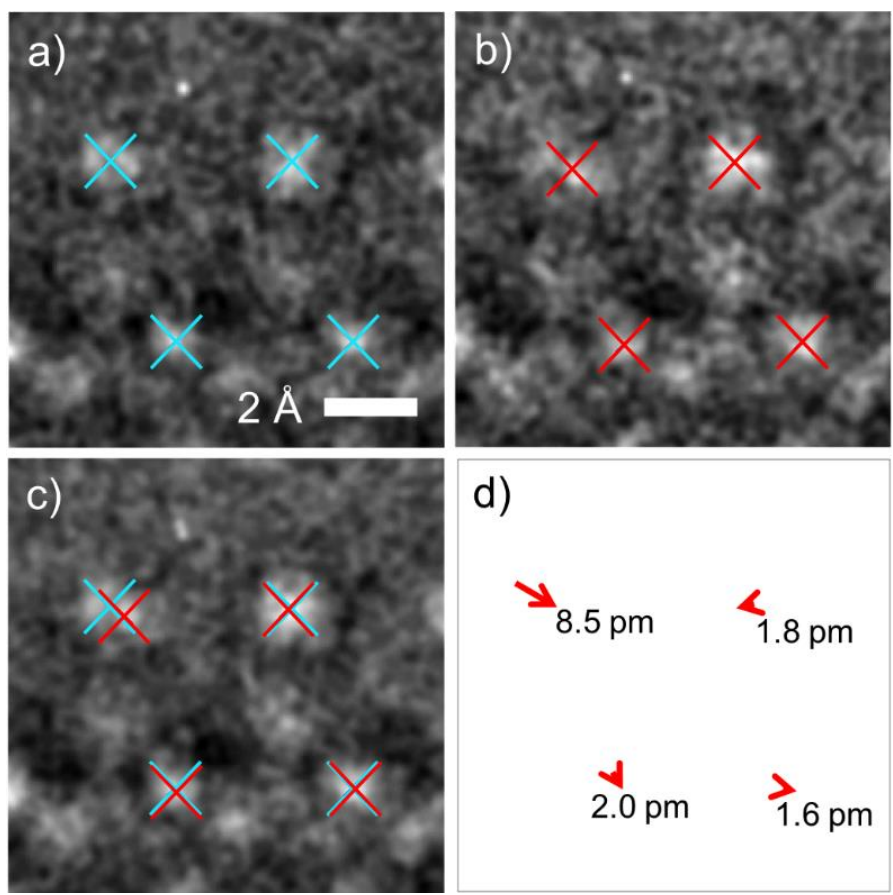

**Figure 3.** a) A frame from an image series of 4 Ce atomic columns on the (111) surface of a $CeO_2$ nanoparticle. The cyan crosses represent the positions of the columns, as found using 2D Gaussian fitting. The frame was acquired with a dose of 500 e$^-$Å$^{-2}$, and the frame was filtered using a Gaussian blur (see Materials and Methods) b) A second frame from the same image series as (a). The red crosses represent the positions of the columns found by using 2D Gaussian fitting. c) A summed image of the frames in (a) and (b) with the cyan and red crosses marking Ce column positions overlaid. This shows that the most significant shift in position between the frames occurred for the top left column. d) A quiver plot showing the direction and magnitude of the shifts in position of the Ce columns between frames (a) and (b).

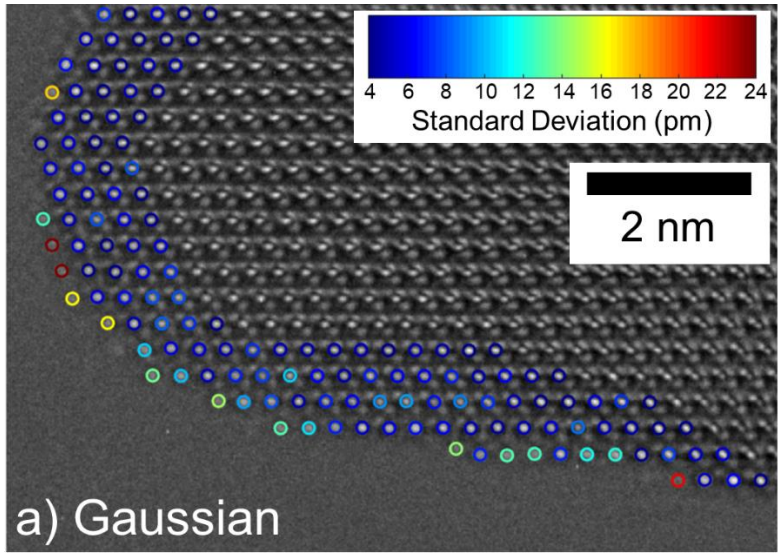
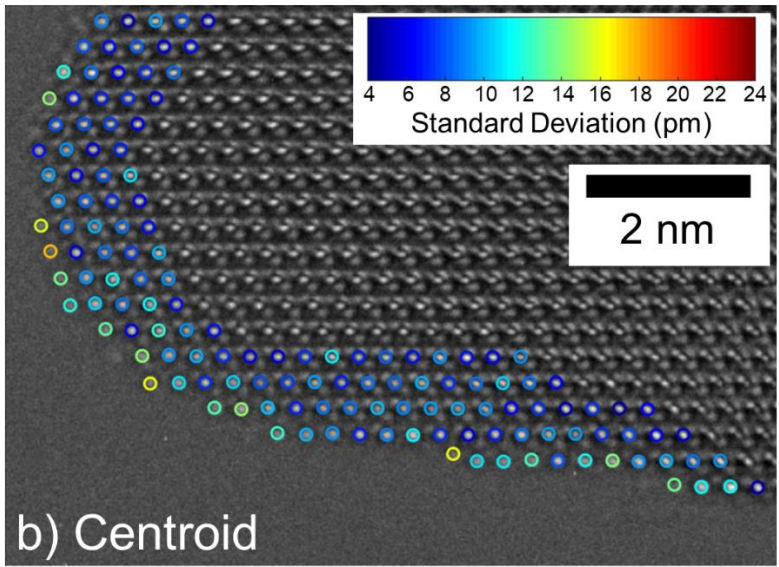
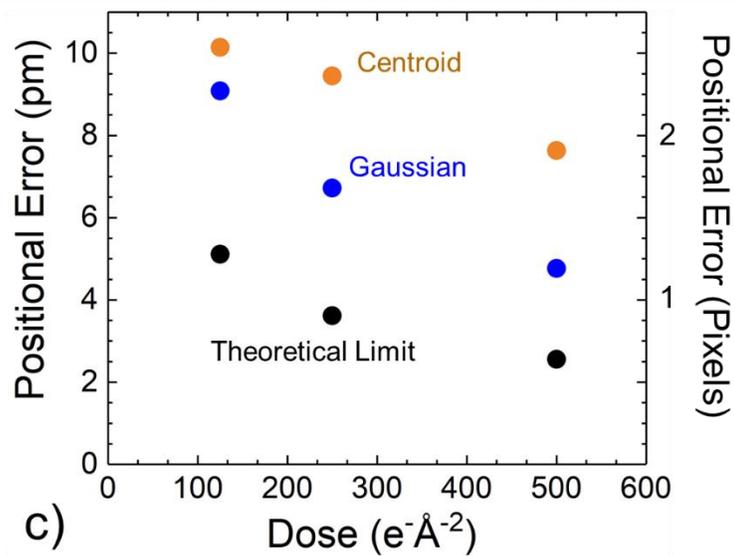

**Figure 4.** a) Summed image of the stepped (111) surface of a $CeO_2$ nanoparticle. Total exposure 1s, sum of 10 frames at 500 e$^-$Å$^{-2}$ per frame. The coloured circles overlaid on the Ce atomic columns indicate the standard deviation of the positions of the Ce columns calculated from Gaussian fitting. A larger standard deviation may imply more frequent, and larger displacements between frames in the image series. The typical standard deviation of a sub-surface column given the dose per frame is ~4.7 pm, which we use to characterize the error on a positional measurement. b) The same image as shown in (a), but with the standard deviation of positions of the Ce columns calculated by Centroid fitting, instead of Gaussian fitting. The results are qualitatively very similar, but the Centroid fitting results in a larger standard deviation of ~7.6 pm for the stationary sub-surface columns, suggesting that the Gaussian fitting method may be slightly more accurate at measuring column positions. c) Error on positional measurement calculated using the Gaussian and Centroid fitting methods compared to a theoretical limit [38.39].

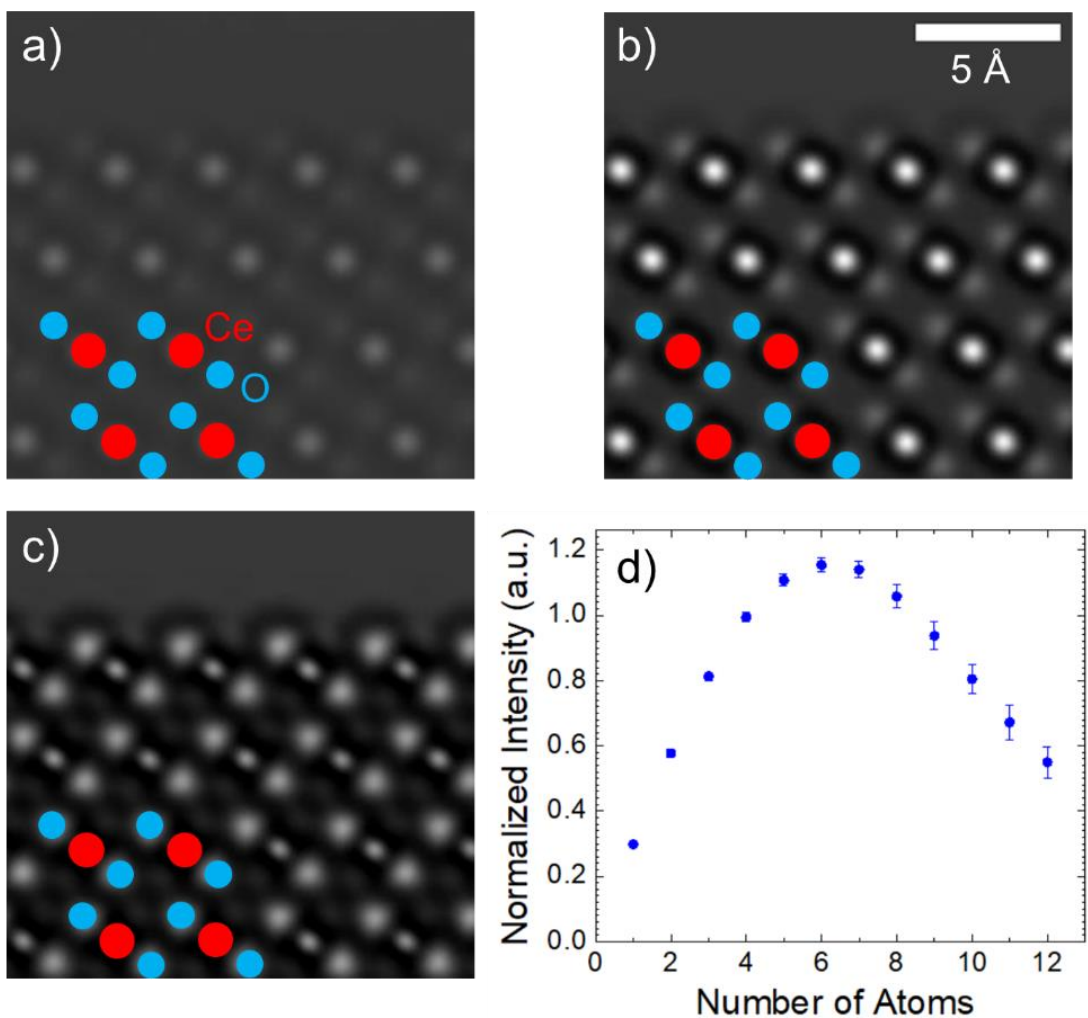

**Figure 5**. Multislice simulations of a (111) $CeO_2$ surface, viewed along the [110] axis at different thicknesses. The imaging conditions for the simulations were chosen to match the conditions of the experimental data in Figure 3 and Figure 4. a) 1 Ce atom per column, b) 5 Ce atoms per column, c) 12 Ce atoms per column. d) Average of measured intensity at 4 different sites. The error bars represent the range of intensities between the sites. For occupancies of less than 5 atoms per column, the relationship between occupancy and intensity is approximately linear, and the difference in intensity between different sites is negligible.

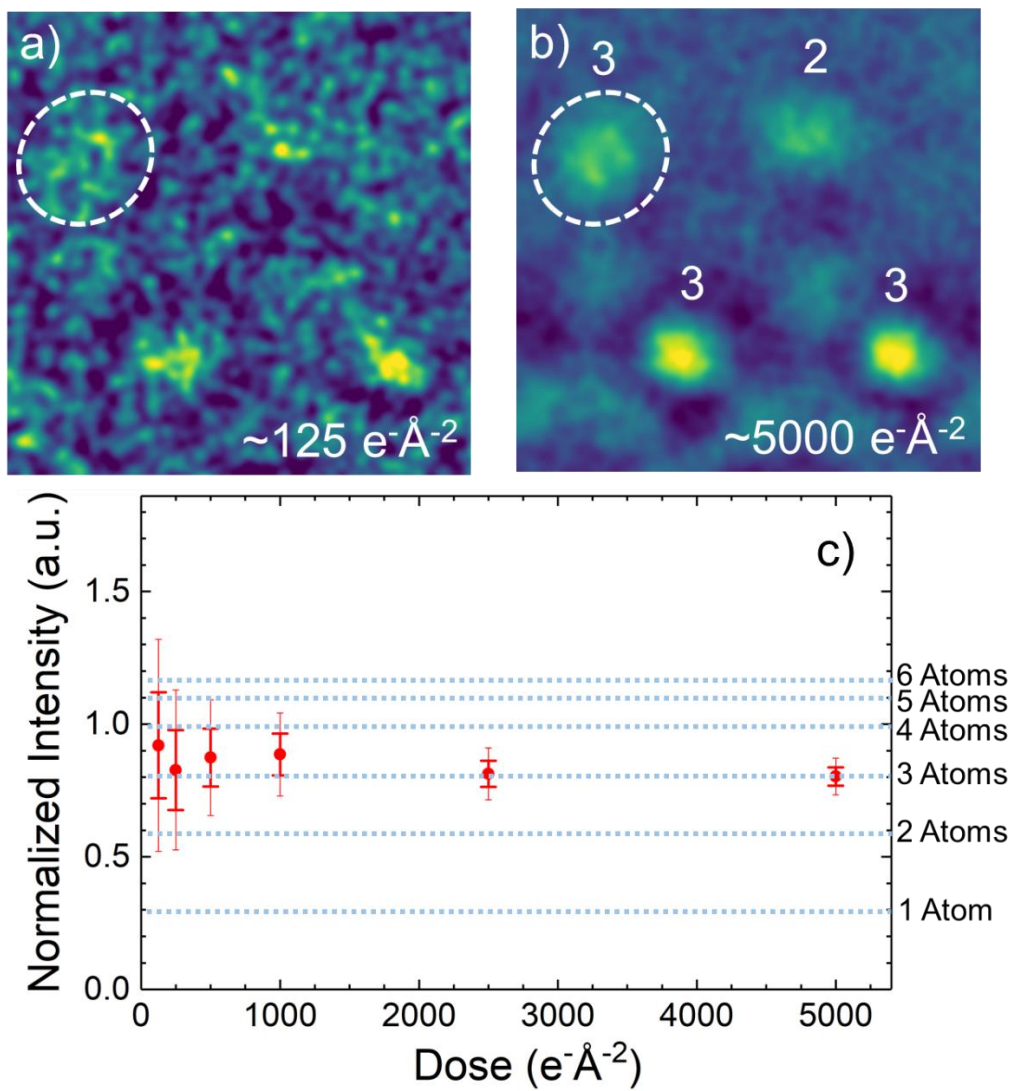

**Figure 6**. a) A frame from an image series of 4 Ce atomic columns on the (111) surface of a $CeO_2$ nanoparticle. This frame was acquired with a dose of ~125 $e^-Å^{-2}$, and the frame was filtered using a Gaussian blur (see Materials and Methods). b) A summed image of 40 frames of the same area of $CeO_2$ as in (a) with a total dose for the image of 5000 $e^-Å^{-2}$, filtered using a Gaussian blur. The numbers above each Ce column indicate the estimated number of Ce atoms in that column based on comparison with multislice simulations c) A graph showing intensity of the atomic column inside the white circle in (a) and (b) calculated for different total doses per image, normalized for comparison with the lookup table in Figure 5d. The error bars are calculated using Equation 5. The bold error bar represents 64% confidence, and the thinner error bar represents 95% confidence. For comparison with the lookup table (Figure 5d), all intensities have been normalized such that the integrated intensity of an area of 1 Å$^2$ in vacuum is equal to 1.

# Supplementary Information

**Exploration of the Impact of Imaging Filters on 2D Gaussian Fitting**

Supplementary Figure 1 shows a comparison of a sample of raw image data, with the same data after Gaussian Blur, Wiener, and TV-L1 Filtering. This sample frame is from the surface of a $CeO_2$ nanoparticle and is notable because the top-right atomic column is in an off-lattice position.

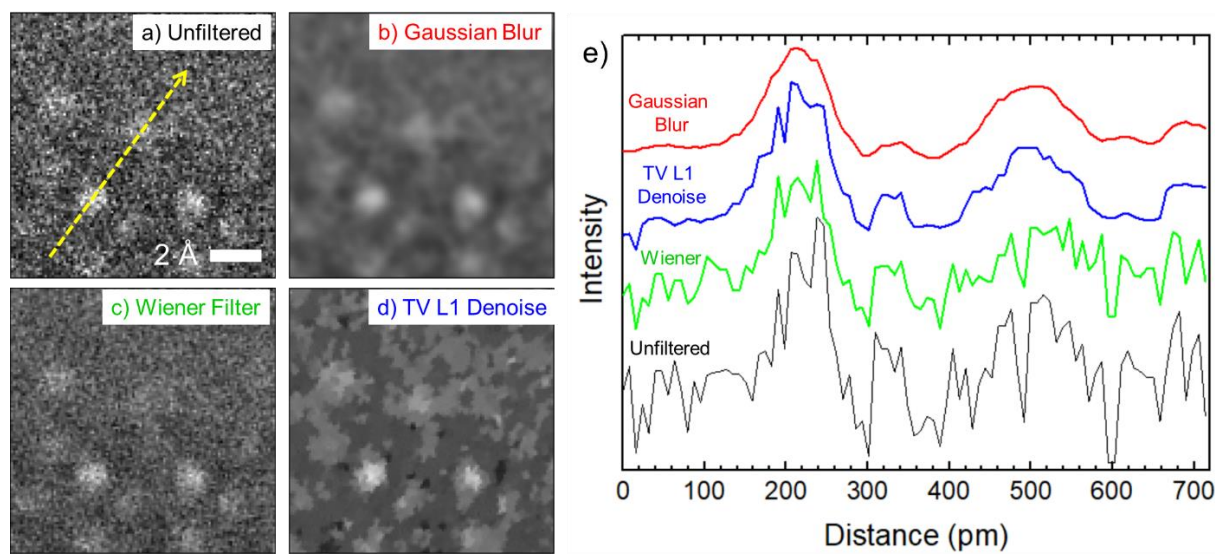

**Supplementary Figure 1**. a) Unfiltered image of the same area of the $CeO_2$ nanoparticle as in Figure 1c with 0.1 s exposure time. b) The result of applying a Gaussian blur filter of radius 2 pixels to (a). c) The result of applying a Weiner filter to (a). d) The result of applying a TV-L1 denoising filter to (a). e) Intensity profiles through each of the images (a) – (d) along the line indicated by the dashed arrow in (a). The Intensity profiles are vertically offset from each other but have not been scaled. This highlights the apparent reduction in noise and smoothing of data due to each of the image filters.

The Gaussian Blur filter is implemented in ImageJ and operates by convolving a 2D image with a 2D circular Gaussian function. A radius of 2 pixels was chosen for the Gaussian blur filtered image, which corresponds to a FWHM of ~4.7 pixels, which is much narrower than a typical atomic column. The Wiener filter was implemented in MATLAB, based on the description given by Robert Kilaas [Ref. 32 in the main manuscript]. Wiener Filtering has been used in TEM imaging to enhance the contrast of crystal lattice on amorphous backgrounds for a number of years [Ref. 33 in the main manuscript]. However, we have observed that by enhancing the contrast of periodic structures, the Wiener filter can also reduce the visibility of defects and can even alter the apparent position of atomic columns. This effect is clear in Supplementary Figure 1c, where the visibility of the off-lattice site atomic column in the top right is clearly reduced relative to the other atomic columns and is investigated further below. Since dynamic changes in surface structure may result in displacements of atomic columns from lattice sites, as in Figure 1, this suggests that the Wiener filter may be unsuitable for locating the precise, sub-pixel level position of atomic columns in temporally resolved structural analysis. The TV-L1 denoising filter attempts to generate an image that is similar to the raw data, but with a much lower variance, in an effort to preserve important image features such as edges whilst removing noise [Ref. 34 in the main manuscript]. We found that the TV-L1 filter generally produced images that appeared reasonable when viewed at lower magnifications, but when zooming in to inspect potential shifts in atomic positions on the order of a few pixels, the filtered images appeared patchy, and difficult to interpret (e.g. Supplementary Figure 1d). The effects of each of the three filters on the Gaussian fitting are explored in Supplementary Figure 2 and Supplementary Table 1 below.

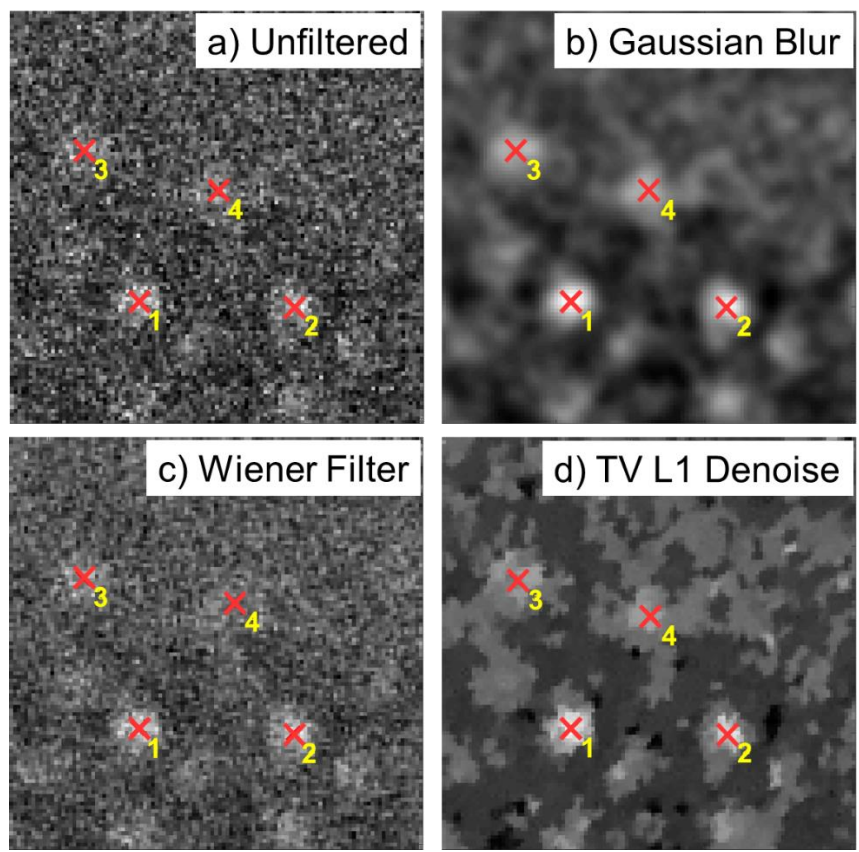

**Supplementary Figure 2.** Positions of atomic columns identified by 2D Gaussian fitting for a) Unfiltered data b) Gaussian blur filtered data c) Wiener filtered data d) TV-L1 filtered data. These positions are quantified and compared in Supplementary Table 1, along with the intensities and $R^2$ values of the fitted Gaussians.

| Filter Type | Atom ID No. | $R^2$ | X (Pixels) | Y (Pixels) | X Difference (Pixels) | Y Difference (Pixels) | Intensity (arb. units) | Intensity Difference (%) |
|---|---|---|---|---|---|---|---|---|
| **Unfiltered** | 1 | **0.848** | 43.95 | 78.48 | **N/A** | **N/A** | 273.67 | **N/A** |
| | 2 | **0.846** | 86.06 | 79.85 | **N/A** | **N/A** | 234.74 | **N/A** |
| | 3 | **0.886** | 29.38 | 38.06 | **N/A** | **N/A** | 257.65 | **N/A** |
| | 4 | **0.858** | 64.98 | 48.92 | **N/A** | **N/A** | 299.73 | **N/A** |
| | | | | | | | | |
| **Gaussian Blur Filter** | 1 | **0.982** | 43.83 | 78.50 | **-0.12** | **0.02** | 251.98 | **-7.93** |
| | 2 | **0.980** | 85.76 | 80.16 | **-0.30** | **0.31** | 223.22 | **-4.90** |
| | 3 | **0.993** | 29.33 | 38.09 | **-0.05** | **0.03** | 240.34 | **-6.71** |
| | 4 | **0.990** | 65.26 | 48.62 | **0.28** | **-0.30** | 301.42 | **0.56** |
| | | | | | | | | |
| **Wiener Filter** | 1 | **0.942** | 43.89 | 78.97 | **-0.06** | **0.49** | 252.61 | **-7.70** |
| | 2 | **0.944** | 85.59 | 80.08 | **-0.47** | **0.23** | 226.75 | **-3.40** |
| | 3 | **0.957** | 29.19 | 38.69 | **-0.19** | **0.63** | 192.50 | **-25.29** |
| | 4 | **0.949** | 69.82 | 44.87 | **4.84** | **-4.04** | 310.30 | **3.53** |
| | | | | | | | | |
| **TV L1 Filter** | 1 | **0.965** | 43.86 | 78.59 | **-0.08** | **0.11** | 313.42 | **14.52** |
| | 2 | **0.964** | 85.98 | 80.27 | **-0.07** | **0.42** | 221.34 | **-5.71** |
| | 3 | **0.977** | 29.48 | 39.19 | **0.10** | **1.13** | 296.89 | **15.23** |
| | 4 | **0.966** | 65.54 | 48.38 | **0.55** | **-0.53** | 295.93 | **-1.29** |

**Supplementary Table 1.** Positional coordinates, intensities and R2 values associated with 2D elliptical Gaussian fits to the atomic columns shown in the images in Supplementary Figure 2.

Each of the filters tested resulted in an improved $R^2$ value for Gaussian fitting compared to the raw data, which may be due to the fact that all of the filters smoothed the data to some degree. Of the filters tested, the Gaussian blur filter appeared to introduce the fewest image artifacts. Changes in the measured position of columns before and after Gaussian blur filtering are relatively small (0.31 pixels or less in the X and Y directions), and changes in measured intensity are relatively small (< 8%). Differences from the measured intensities and positions in the unfiltered data are greater for the TV L1 and Wiener filters. An important observation is that the measured position of atomic column 4 changes by >4 pixels in both X and Y after applying a Wiener filter, which appears to be due to the fact that the Wiener filter imposes a periodicity on the image. This indicates that the Wiener filter is not appropriate for measuring the positions of columns in off-lattice defect sites. Another advantage of the Gaussian blur filter (implemented in ImageJ) was that it could be applied to an entire image series in less than 1 second, which was faster than the MATLAB implementations of the Wiener Filter and TV-L1 denoising filter available to us. Therefore, we chose to use the Gaussian blur filter to pre-process data prior to atom tracking.

**Error on Atomic Column Position as a Function of Beam Dose per Frame**

| Estimated Dose Per Frame (e$^-$Å$^{-2}$) | Positional Error Centroid (pm) | Positional Error Gaussian (pm) | Theoretical Best Precision |
|---|---|---|---|
| 125 | 10.14 | 9.08 | 5.11 |
| 250 | 9.45 | 6.72 | 3.61 |
| 500 | 7.63 | 4.76 | 2.56 |

**Supplementary Table 2.** Error on position as a function of dose per frame (see also Figure 4c in the main text). These errors were calculated as the average of the standard deviation of the positions of 25 sub-surface Ce columns, which were expected to remain static over the course of the experiment. The range of doses per frame considered is limited from 125 to 500 e$^-$Å$^{-2}$ for statistical reasons, i.e. for doses larger than 500 e$^-$Å$^{-2}$ per frame, there would be fewer than 10 frames over which we could measure standard deviations, meaning that the results may be statistically unreliable. For our dataset, the 2D Gaussian fitting method gave a lower error on position than the centroid method over the range of doses per frame considered. Both methods give a larger error in position than the theoretical expression for the best attainable precision.

**Parameters for Multislice Image Simulations**

High Tension: 300 kV.

Defocus: 7.8 nm.

Defocus Spread: 1 nm.

2-fold Astigmatism: 0 nm.

3-fold Astigmatism: 0 nm.

2$^{nd}$ Order Coma: 0 nm.

3$^{rd}$ Order Spherical Aberration: 30 µm

5$^{th}$ Order Spherical Aberration: 5 mm.

Chromatic Aberration: 1 mm.

**Error on Atomic Column Intensities as a Function of Beam Dose per Frame**

We have considered two methods for estimating the error on measured column intensities. The first method considered is to estimate the error based on the measured intensities assuming Poisson noise, as described in Methods, with results shown in Figure 6. The alternative method is to measure the standard deviation of a series of measurements of intensity of an atomic column. The advantage of using the standard deviation as an estimate of error is that it makes no assumptions about the type of noise in the images. The disadvantage of the using standard deviation as an estimate of error is that genuine changes in atomic column intensity due to dynamic changes in column occupancy may lead to the standard deviation being an overestimate of the error. Such changes in column occupancy may occur due to electron beam damage, or to the migration of weakly bound surface atoms.

When using the standard deviation to estimate the error in intensity, it is also important to consider that the error on measuring the intensity of an atomic column is likely to be a function of the intensity per pixel in the atomic column. For example, if Poisson noise was the primary contribution to noise, the error on intensity in a pixel \would be proportional to the square root of the measured intensity of that pixel. This means that measuring the standard deviation of the intensity of sub-surface columns, where the intensity per pixel is relatively bright, may not yield an accurate estimate of error for the intensity of more diffuse surface columns, where intensity per pixel is relatively low. This is highlighted in Supplementary Figure 3a, which shows the standard deviation in intensity measurements for different Ce columns in a $CeO_2$ nanoparticle measured over 10 frames at 500 $e^-Å^{-2}$ per frame. Sub-surface columns, which typically exhibit a greater intensity per pixel, but occupy fewer pixels in area, show a smaller standard deviation in intensity than surface columns, which typically exhibit a lower intensity per pixel, but occupy a greater number of pixels in area. It is also possible that some changes in column occupancy during the image series may contribute to the large standard deviation of intensity observed for some of the surface columns. We do not observe a decrease in intensity for any of these columns over the image series, so removal of Ce atoms by electron beam damage is unlikely, although migration of weakly bound surface adatoms is not ruled out.

In order to directly compare the estimate of error on intensity due to standard deviation and our Poisson noise analysis, we chose to investigate the standard deviation in intensity of the same column that we performed a Poisson noise analysis on in Figure 6 in the main manuscript. As shown in Figure 6c, the measured intensity of this atomic column corresponded very precisely to 3 Ce atoms at a beam dose of 2500-5000 $e^-Å^{-2}$ per frame. This gives us some confidence that the occupancy of this atomic column did not change during the acquisition of data (in which case the measured intensity of the column would have been expected to fall between values that correspond to integer occupancy). The standard deviation of the intensity of this column should therefore be unaffected by a change in occupancy, and should give a more reasonable estimate of error. Supplementary Figure 3b shows an image, highlighting the atomic column used for measurements. Supplementary Figure 3c compares error derived from Poisson noise analysis to the standard deviation. The standard deviation tends to be larger than the error assuming Poisson noise, but seems to converge towards the error assuming Poisson noise as dose per frame increases.

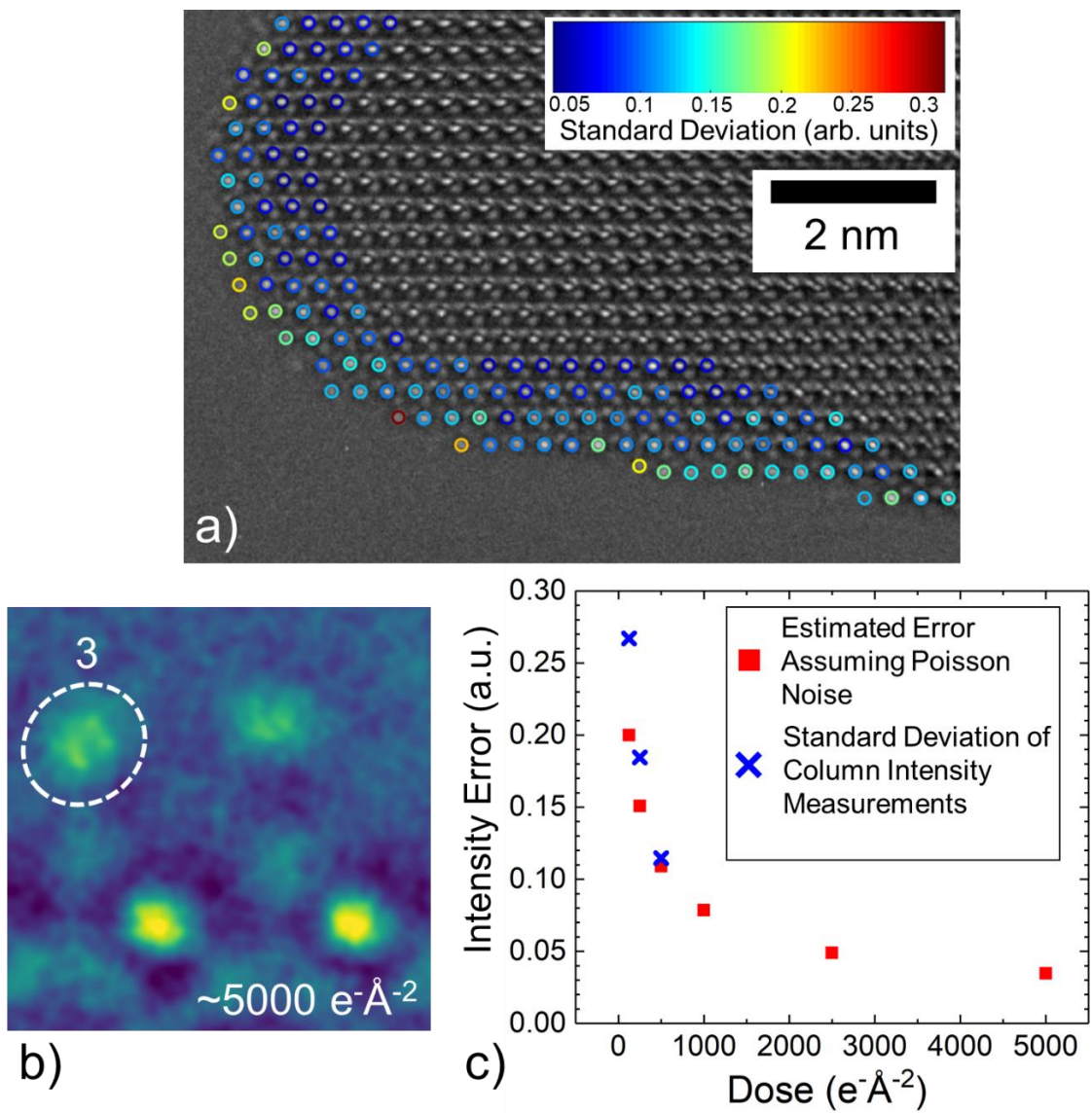

**Supplementary Figure 3.** a) Summed image of the stepped (111) surface of a $CeO_2$ nanoparticle. Total exposure 1s, sum of 10 frames at 500 e⁻Å⁻² per frame. The coloured circles overlaid on the Ce atomic columns indicate the standard deviation of the intensity of the Ce columns calculated from Gaussian fitting. b) A summed image of 40 frames of the same area of $CeO_2$ as in Figure 6b with a total dose for the image of 5000 e⁻Å⁻², filtered using a Gaussian blur. The numbers above each Ce column indicate the estimated number of Ce atoms in that column based on comparison with multislice simulations. c) Estimates for the error on the measured intensity as a function of dose for the top left atomic column in (a) using two different methods. The red squares indicate error estimated assuming Poisson noise (data from Figure 6c). The blue crosses indicate the standard deviation of multiple intensity measurements on the top left column from successive frames in an image series. The range of doses per frame considered is limited from 125 to 500 e$^-$Å$^{-2}$ for statistical reasons, i.e. for doses larger than 500 e$^-$Å$^{-2}$ per frame, there would be fewer than 10 frames over which we could measure standard deviations, meaning that the results may be statistically unreliable.